# Circular birefringence in crystal optics

R J Potton [a]

*Joule Physics Laboratory,*

*School of Computing, Science and Engineering,*

*Materials and Physics Research Centre,*

*University of Salford,*

*Greater Manchester M5 4WT, UK.*

**Abstract**

In crystal optics the special status of the rest frame of the crystal means that space-time symmetry is less restrictive of electrodynamic phenomena than it is of static electromagnetic effects. A relativistic justification for this claim is provided and its consequences for the analysis of optical activity are explored. The discrete space-time symmetries P and T that lead to classification of static property tensors of crystals as polar or axial, time-invariant (-i) or time-change (-c) are shown to be connected by orientation considerations. The connection finds expression in the dynamic phenomenon of gyrotropy in certain, symmetry determined, crystal classes. In particular, the degeneracies of forward and backward waves in optically active crystals arise from the covariance of the wave equation under space-time reversal.

[a] Electronic mail: R.J.Potton@salford.ac.uk

# 1. Introduction

To account for optical activity in terms of the dielectric response in crystal optics is more difficult than might reasonably be expected [1]. Consequently, recourse is typically had to a phenomenological account. In the simplest cases the normal modes are assumed to be circularly polarized so that forward and backward waves of the same handedness are degenerate. If this is so, then the circular birefringence can be expanded in even powers of the direction cosines of the wave normal [2]. The leading terms in the expansion suggest that optical activity is an allowed effect in the crystal classes having second rank property tensors with non-vanishing symmetrical, axial parts.

To reconcile the phenomenological approach with a local theory of dielectric response one needs to examine carefully the symmetry restriction on permittivity tensor elements. In doing this one must recognise the special status of the rest frame of the crystal and its connection with time-reversal symmetry [3]. The upshot of this examination is that one should focus attention on subgroups of the proper Poincaré group, containing space-time inversion PT but not P and T separately, rather than on subgroups of the full Poincaré group.

In section 2, Maxwell's equations for a polarisable medium are set out, following Berreman, in a form that lends itself to detailed analysis of the phenomenon in question. Electro-magnetic constitutive equations for the *dynamic* response in gyrotropic media are introduced in section 3 by way of extension of Birss's formulation [4] of similar relations for *static* effects. A justification for the inclusion of additional elements in the dynamic permittivity, permeability and conductivity tensors is provided in an appendix. In section 4 wave equations that exhibit the PT symmetry responsible for the degeneracy of forward and backward waves of the same handedness are developed. Section 5 contrasts the present approach, based on the space-time symmetry of the permittivity tensor, with phenomenological treatments that expand circular birefringence, expressed as a pseudo-scalar, in direction cosines of the wave normal. The corresponding listings of gyrotropic crystal classes do not exactly match providing the opportunity of experimentally testing what is set out here.

# 2. Maxwell's equations in a polarizable medium

The aim is to explain the interaction of plane polarized components of a circularly polarized wave in terms of the constitutive relations of a less than fully symmetric medium. To facilitate this it is expedient to employ the 4×4 matrix formulation developed by Teitler [5] and Berreman [6].

The Ampere-Maxwell law:

$$\text{curl } \mathbf{H} = \mathbf{J}_f + \frac{\partial \mathbf{D}}{\partial t} \tag{2.1}$$

and the constitutive relations:

$$\mathbf{D} = \boldsymbol{\varepsilon}\varepsilon_0 \mathbf{E} \tag{2.2}$$

and

$$\mathbf{H} = \frac{\boldsymbol{\mu}^{-1}}{\mu_0}\mathbf{B} \tag{2.3}$$

and

$$\mathbf{J}_f = \boldsymbol{\sigma}\mathbf{E} \tag{2.4}$$

are the starting points for the analysis that follows.

Notwithstanding the conventional conflating of conductivity and dielectric response in a single, complex dielectric tensor, the form of (2.1) makes it natural to keep separate the conductivity and dielectric displacement contributions to curl $\mathbf{H}$. They act in quadrature and dissipation is associated with $\boldsymbol{\sigma}$ and not with $\boldsymbol{\varepsilon}$.

Linear and homogeneous but anisotropic modelling of the crystalline medium is adequate for illustrating the points to be made. Maxwell's curl equations take the form:

$$\partial_r \times \mathbf{E} + \partial_t \mathbf{B} = 0 \tag{2.5}$$

$$\partial_r \times \frac{\boldsymbol{\mu}^{-1}}{\mu_0}\mathbf{B} - \partial_t(\varepsilon_0 \boldsymbol{\varepsilon})\mathbf{E} - \boldsymbol{\sigma}\mathbf{E} = 0 \tag{2.6}$$

To illustrate the matrix method, the medium is initially taken to be dielectric and isotropic:

$$\boldsymbol{\sigma} = \mathbf{0} \tag{2.7}$$

$$\boldsymbol{\varepsilon} = \varepsilon \mathbf{I} \tag{2.8}$$

$$\boldsymbol{\mu} = \mu \mathbf{I} \tag{2.9}$$

and a refractive index, $n$, is defined by $n/c = \sqrt{\varepsilon_0 \mu \mu_0}$. When particularized to plane waves travelling along the z-axis Maxwell's curl equations take the form:

$$\partial_z \frac{n}{c} E_x = -\frac{n}{c}\partial_t B_y \tag{2.10}$$

$$\partial_z B_y = -\frac{n}{c}\partial_t \frac{n}{c} E_x \tag{2.11}$$

$$\partial_z \frac{n}{c} E_y = \frac{n}{c} \partial_t B_x \tag{2.12}$$

$$\partial_z B_x = \frac{n}{c} \partial_t \frac{n}{c} E_y \tag{2.13}$$

or as a matrix equation with basis $\left\{ \frac{n}{c} E_x, B_y, \frac{n}{c} E_y, B_x \right\}$:

$$\begin{bmatrix} \partial_z & (n/c)\partial_t & 0 & 0 \\ (n/c)\partial_t & \partial_z & 0 & 0 \\ 0 & 0 & \partial_z & -(n/c)\partial_t \\ 0 & 0 & -(n/c)\partial_t & \partial_z \end{bmatrix} \begin{bmatrix} (n/c)E_x \\ B_y \\ (n/c)E_y \\ B_x \end{bmatrix} = 0 \tag{2.14}$$

The equation is diagonal in the basis $\frac{1}{\sqrt{2}} \left\{ \frac{n}{c} E_x + B_y, \frac{n}{c} E_y + B_x, \frac{n}{c} E_y - B_x, \frac{n}{c} E_x - B_y \right\}$:

$$\begin{bmatrix} \partial_z + \frac{n}{c}\partial_t & 0 & 0 & 0 \\ 0 & \partial_z - \frac{n}{c}\partial_t & 0 & 0 \\ 0 & 0 & \partial_z + \frac{n}{c}\partial_t & 0 \\ 0 & 0 & 0 & \partial_z - \frac{n}{c}\partial_t \end{bmatrix} \begin{bmatrix} \frac{n}{c} E_x + B_y \\ \frac{n}{c} E_y + B_x \\ \frac{n}{c} E_y - B_x \\ \frac{n}{c} E_x - B_y \end{bmatrix} = 0 \tag{2.15}$$

The eigenmodes are respectively forward *x*-polarized, backward *y*-polarized, forward *y*-polarized and backward *x*-polarized with dispersion relations $k_z = +n\omega/c$, $k_z = -n\omega/c$, $k_z = +n\omega/c$ and $k_z = -n\omega/c$ respectively as may be verified by consideration of the Poynting vectors of the respective field superpositions.

In an isotropic medium, circularly polarized modes could equally serve as a basis (the helicity basis).

## 3. Constitutive relations for harmonic fields

The discrete symmetries of space and time are variously denoted by P and T or I and $I_t$ [7] defined as follows:

$$\mathrm{P}: (x, y, z) \mapsto (-x, -y, -z) \tag{3.1}$$

$$\mathrm{T}: t \mapsto -t \tag{3.2}$$

are elements of the full inhomogeneous Lorentz (full Poincaré) group [8].

When applied to *static* fields the dielectric constitutive relation (2.2) relates fields that have definite transformation properties under discrete space-time symmetries [4]. Both **E** and **D** are, in fact, odd under P (polar) and even under T (-i) [9] [†]. Consequently, $\varepsilon$ is a polar-i tensor. The polar-i designation is the one used by Birss. It is germane to ask whether any discrete space-time symmetry persists generally (that is to say, independently of any particular crystalline point symmetry) for dynamic fields (i.e. waves) in crystalline condensed matter. In pursuance of this it is prudent to examine how the application of full Poincaré group symmetry operations must differ when light propagates in a crystalline medium. Appendix A shows that, in a crystalline medium, an association can be made between oriented spatial entities and ordering in time. This association is invariant under combined space and time inversion which is a symmetry operation of the proper Poincaré group:

$$\mathrm{PT}: (t,x,y,z) \mapsto (-t,-x,-y,-z) \tag{3.1}$$

but not under separate space inversion, P, and time inversion, T, which are not. Consequently the ruling symmetry in crystal optics is found to be the proper Poincaré group depleted of all Lorentzian boosts and of all but a discrete set of proper and improper spatial rotations. Matrix and tensor representation of the Poincaré group give rise to complex amplitudes which are affected by time re-ordering because of the anti-unitary character of space-time reversal.

It is well accepted that time-reversal:

$$\begin{aligned}\Theta: t &\mapsto -t \\ \mathrm{c} &\mapsto \mathrm{c}*\end{aligned} \tag{3.2}$$

where c is any complex amplitude, is a prevalent symmetry (for example of Schrodinger's wave equation) [10] whereas time inversion:

$$\mathrm{T}: t \mapsto -t \tag{3.3}$$

is not [7]. By analogy with the former it is possible to introduce space-time reversal:

---

[†] It will be seen shortly that linear boosts are excluded from the set of symmetry operations so fields can be treated as vectors rather than components of second rank tensors.

R : $t \mapsto t' = -t$

$\boldsymbol{\alpha} \mapsto \boldsymbol{\alpha}' = \boldsymbol{\alpha}$   even rank polar · i ; odd rank axial · i ; even rank axial · c ; odd rank polar · c

$\boldsymbol{\alpha} \mapsto \boldsymbol{\alpha}' = -\boldsymbol{\alpha}$   even rank polar · c ; odd rank axial · c ; even rank axial · i ; odd rank polar · i

c ↦ c *      (3.4)

where $\boldsymbol{\alpha}$ is a tensor or a pseudo(axial)-tensor and c is a complex amplitude.

In section 4 the constitutive equation:

$$\mathbf{D} = \boldsymbol{\varepsilon}\mathbf{E} \tag{3.5}$$

is the subject of attention. In accordance with the discussion in Appendix A the terms in this equation are taken to be classifiable under space-time reversal, R. **D** and **E** are odd [9], hence, $\boldsymbol{\varepsilon}$ is even. However, in addition to the usual polar-i part, represented by a real, symmetric second rank tensor, $\boldsymbol{\varepsilon}$ has an axial-c part. Its tensor representative is anti-symmetric and, in non-magnetic crystal classes, pure imaginary. This part of $\boldsymbol{\varepsilon}$ is non-zero when fields are not static and the sequence in which *x*- and *y*-components change matters i.e. in the presence of gyrotropy.

The essence of the argument in section 4 is that the degeneracy of circularly polarized waves has its origin in the space-time reversal covariance of the wave equation.

## 4. Circular birefringence

If the permittivity tensor $\boldsymbol{\varepsilon}$ has off-diagonal elements $\varepsilon_{xy}$ and $\varepsilon_{yx}$ equations (2.11) and (2.13) become:

$$\partial_z B_y = -\frac{n}{c}\partial_t \frac{n}{c}E_x - \frac{\varepsilon_{xy}}{\varepsilon}\frac{n}{c}\partial_t \frac{n}{c}E_y \tag{4.1}$$

$$\partial_z B_x = \frac{n}{c}\partial_t \frac{n}{c}E_y + \frac{\varepsilon_{yx}}{\varepsilon}\frac{n}{c}\partial_t \frac{n}{c}E_x \tag{4.2}$$

and the matrix equation:

$$\begin{bmatrix} \partial_z & (n/c)\partial_t & 0 & 0 \\ (n/c)\partial_t & \partial_z & (\varepsilon_{xy}/\varepsilon)(n/c)\partial_t & 0 \\ 0 & 0 & \partial_z & -(n/c)\partial_t \\ -(\varepsilon_{yx}/\varepsilon)(n/c)\partial_t & 0 & -(n/c)\partial_t & \partial_z \end{bmatrix} \begin{bmatrix} (n/c)E_x \\ B_y \\ (n/c)E_y \\ B_x \end{bmatrix} = 0 \tag{4.3}$$

in the original basis and:

$$\begin{bmatrix} \partial_z + \dfrac{n}{c}\partial_t & \dfrac{\varepsilon_{xy}}{2\varepsilon}\dfrac{n}{c}\partial_t & \dfrac{\varepsilon_{xy}}{2\varepsilon}\dfrac{n}{c}\partial_t & 0 \\ -\dfrac{\varepsilon_{yx}}{2\varepsilon}\dfrac{n}{c}\partial_t & \partial_z - \dfrac{n}{c}\partial_t & 0 & -\dfrac{\varepsilon_{yx}}{2\varepsilon}\dfrac{n}{c}\partial_t \\ \dfrac{\varepsilon_{yx}}{2\varepsilon}\dfrac{n}{c}\partial_t & 0 & \partial_z + \dfrac{n}{c}\partial_t & \dfrac{\varepsilon_{yx}}{2\varepsilon}\dfrac{n}{c}\partial_t \\ 0 & -\dfrac{\varepsilon_{xy}}{2\varepsilon}\dfrac{n}{c}\partial_t & -\dfrac{\varepsilon_{xy}}{2\varepsilon}\dfrac{n}{c}\partial_t & \partial_z - \dfrac{n}{c}\partial_t \end{bmatrix} \begin{bmatrix} \dfrac{n}{c}E_x + B_y \\ \dfrac{n}{c}E_y + B_x \\ \dfrac{n}{c}E_y - B_x \\ \dfrac{n}{c}E_x - B_y \end{bmatrix} = 0 \qquad (4.4)$$

in the basis of linearly polarized forward and backward modes.

The axial-c part of the permittivity tensor yields a pure imaginary and antisymmetric $\boldsymbol{\varepsilon}$. The antisymmetry of $\boldsymbol{\varepsilon}$ arises because of the dependence of dielectric response on the sequence in which changes in different components of the electric field occur. This is a feature of PT symmetry which is appropriate for gyrotropic media. In non-magnetic crystal classes time reversal is a symmetry that stands on its own (in contrast to magnetic classes in which it occurs only in combination with certain point symmetries). For non-magnetic classes property tensors are either even in time (pure real) or odd in time (pure imaginary).

Given the above considerations (4.4) is put into nearly block diagonal form by the unitary transformation $\mathbf{U} = \dfrac{1}{\sqrt{2}}\begin{bmatrix} \mathbf{I} & +i\mathbf{I} \\ \mathbf{I} & -i\mathbf{I} \end{bmatrix}$ where $\mathbf{I}$ is the $2\times 2$ unit matrix:

$$\begin{bmatrix} \partial_z + \left(1 - i\dfrac{\varepsilon_{xy}}{2\varepsilon}\right)\dfrac{n}{c}\partial_t & 0 & 0 & \dfrac{\varepsilon_{xy} - \varepsilon_{yx}}{4\varepsilon}\dfrac{n}{c}\partial_t \\ 0 & \partial_z - \left(1 - i\dfrac{\varepsilon_{yx}}{2\varepsilon}\right)\dfrac{n}{c}\partial_t & \dfrac{\varepsilon_{xy} - \varepsilon_{yx}}{4\varepsilon}\dfrac{n}{c}\partial_t & 0 \\ 0 & \dfrac{\varepsilon_{xy} - \varepsilon_{yx}}{4\varepsilon}\dfrac{n}{c}\partial_t & \partial_z + \left(1 + i\dfrac{\varepsilon_{xy}}{2\varepsilon}\right)\dfrac{n}{c}\partial_t & 0 \\ \dfrac{\varepsilon_{xy} - \varepsilon_{yx}}{4\varepsilon}\dfrac{n}{c}\partial_t & 0 & 0 & \partial_z - \left(1 + i\dfrac{\varepsilon_{yxy}}{2\varepsilon}\right)\dfrac{n}{c}\partial_t \end{bmatrix}$$

(4.5)

the basis being: forward right, $\dfrac{n}{c}E_x + B_y + i\left(\dfrac{n}{c}E_y - B_x\right)$, backward right, $\dfrac{n}{c}E_y + B_x + i\left(\dfrac{n}{c}E_x - B_y\right)$, forward left, $\dfrac{n}{c}E_x + B_y - i\left(\dfrac{n}{c}E_y - B_x\right)$ and backward left, $\dfrac{n}{c}E_y + B_x - i\left(\dfrac{n}{c}E_x - B_y\right)$, waves. The first two basis functions (like the second two) are degenerate by virtue of the invariance under space-time reversal of the factor that precedes $\pm \frac{n}{c} \partial_t$ in the first (second) two rows of the matrix. By way of explanation $\varepsilon$, in non-magnetic classes, is a pure imaginary antisymmetric axial tensor so that:

$$R : i\varepsilon_{xy} \mapsto -i\varepsilon_{xy} = i\varepsilon_{yx} \quad (4.6)$$

The space-time reversal, R, covariance of the wave equations for forward and backward right-handed waves follows from the identification of $\partial_z$ as first rank axial-i and of $\frac{n}{c}\partial_t$ as zeroth rank polar-c so that:

$$R : \partial_z \mapsto \partial_z' = \partial_z \quad (4.7)$$

$$R : \frac{n}{c}\partial_t \mapsto \frac{n}{c}\partial_t' = -\frac{n}{c}\partial_t. \quad (4.8)$$

Moreover the coefficient, a, of $\frac{n}{c}\partial_t$, is R invariant:

$$R : a = \left(1 - i\frac{\varepsilon_{xy}}{2\varepsilon}\right) \mapsto a' = \left(1 - i\frac{\varepsilon_{yx}}{2\varepsilon}\right) = a \quad (4.9)$$

It is to this covariance that the degeneracy of waves of the same handedness is attributable.

The off-diagonal terms in (4.5) suggest that diagonalization might yield superpositions of forward right circular and backward left circular waves and vice-versa. It is straightforward to investigate such superpositions using Jones calculus [11]. They are strange linear polarized standing waves with nodes spaced by $\pi/k$ but in which the time dependence of the fields is phase modulated with position and time. Fortunately we are rescued from such monsters by a selection rule. The matrix elements in question occur in positions that could only possibly admix forward and backward waves of opposite handedness. However, circular birefringence means that translational symmetry precludes any such admixture.

The crystal classes expected to exhibit circular birefringence on the basis of this analysis are listed in the next section and comparison is made with phenomenological analyses employing second and higher rank tensors.

## 5. Predictions of gyrotropic classes

Tables 2 and 4 of reference [4] show that the crystal classes 1, 2, 222, 4, 4mm, 3, 3m, 6 and 6mm have non-vanishing second rank antisymmetric axial-c tensors. On the basis of the analysis of section 4 they are therefore expected to exhibit circular birefringence. The conventional phenomenological theory of birefringence [2] introduces a pseudoscalar, *G*, related to the changes in ordinary and extraordinary refractive indices brought about by gyrotropy. It is then supposed that *G* can be expanded using a symmetric second rank axial tensor, $g_{ij}$, in direction cosines of the wave normal:

$$G = g_{ij}\alpha_i\alpha_j \tag{6.1}$$

For waves in an isotropic medium it amounts to:

$$G = \pm\frac{\varepsilon_R - \varepsilon_L}{2}. \tag{6.2}$$

Of the classes mentioned above, three, 4mm, 3m and 6mm, have vanishing symmetric second rank axial tensors and are not expected from the phenomenology as presented. This is not a necessarily a serious problem since a more precise phenomenological expression is:

$$G = g_{ij}\alpha_i\alpha_j + h_{ijkl}\alpha_i\alpha_j\alpha_k\alpha_l \tag{6.3}$$

and fourth rank axial tensors are non-vanishing in classes 4mm, 3m and 6mm. (The expansion of *G* as a polynomial of even degree in the direction cosines enshrines the assumption of reciprocal propagation of similarly circularly polarized waves.)

On the other hand the classes m, mm2, $\bar{4}$, 422, $\bar{4}$2m, 32, 622, 23 and 432 predicted as optically active by phenomenology are not anticipated as such from section 4. The implication of this is that dielectric response of higher tensor rank is a possible cause of the effect [††]. This leads naturally to the other school of thought [13] which is that gyrotropy is an intrinsically nonlocal phenomenon and that the dielectric properties require at least third rank tensors for their expression:

$$\varepsilon_{ij}^{-1}(\omega,k) = \varepsilon_{ij}^{-1}(\omega) + in\frac{\omega}{c}\delta_{ijl}s_l. \tag{6.4}$$

One feature of this approach is that it causes attention to be focused on non-symmorphic space-group elements in the form of screw axes [1].

Three non-enantiomorphous classes have attracted attention in connection with optical activity; m, mm2 and $\bar{4}$2m [12]. Given that $\bar{4}$ is in fact enantiomorphous the claim that optical activity may be allowed in a crystal, the symmorphic space group of which has a mirror plane, rests on a single case, namely $\bar{4}$2m. The absence of enantiomorphs may in fact be quite a good guide to the absence optical activity in crystals with symmorphic space groups since it does not depend on arbitrarily truncated series expansions. The non-enantiomorphous classes m, mm2 and $\bar{4}$2m are not predicted as optical active by the analysis of section 4.

†† Another possibility is that the axial-c part of the magnetic permeability or the axial-i part of the conductivity tensor is responsible.

## 6. Further work

The association of time reversal with certain point symmetry operations that exists in magnetic groups [4] must, by Neumann's principle, influence the allowed elements of the dielectric tensor $\varepsilon$. Particularization of symmetry restrictions in these cases will require further study.

The generalization of equation (4.2) to $\sigma \neq 0$ suggests a way of investigating CPT symmetry [3]. $\sigma$ connects odd in C current density with even in C electric field **E** and is therefore odd in C and presumably, therefore, also odd in PT [9]. This connection seems to result from "internal" source terms that are present in conductive media but not in dielectrics.

## 7. Conclusion

It has been shown that, for certain crystal classes, a wave equation exhibiting covariance under space-time reversal (but not under separate space inversion or time reversal) can account for the birefringence and degeneracies of circularly polarized waves.

**Appendix A  A connection between space and time orientations in crystal optics**

This appendix shows that in crystal optics PT symmetry supersedes separate P and T symmetries. There are significant implications for the derivation of crystal symmetry forbidden effects. The argument adduced hinges on the existence of crystalline space-time structures that are invariant under PT but not under P or T separately.

The setting of the discussion is Minkowski space-time and the various homogeneous Lorentz and inhomogeneous Lorentz (Poincaré) groups [8] under which its structures are invariant or covariant. It is first necessary to recognise that crystal optics does not admit Lorentzian linear boosts because crystal symmetry restrictions on property tensors are meaningless except in the rest frame of the crystal by virtue of the phenomenon of Lorentz contraction. A cubic crystal is no longer cubic in a boosted frame. Crystal optics necessarily resides in the rest frame of the crystal. Moreover, this frame is related to the laboratory frame by a unique linear boost. Nevertheless, frame transformations, including finite static rotations and crucially space-time translations remain under which all equations must be covariant.

A significant part of the space-time structure that remains in the crystal rest frame relates to causality. The crystal space-time may be regarded as a partially ordered set (poset) of events by virtue of the fact that, if event B lies within the future light cone of A, we may write $B \leq A$. The partial order relation $\leq$ may in turn be used to define a least upper bound (l.u.b.s, $\cup$) for pairs of events. The l.u.b. of C and D may be defined as the most recent event, the future light cone of which contains both C and D. This definition gives precision to the everyday concept of proximate cause and might be used elsewhere to extend discussion of causation, which is often approached by way of Boolean algebra, to the relativistic domain. However, the properties of this algebra are different in crucial ways from Boolean algebra. Differences emerge when different dimensions of space-time are considered. In 2+1 and 3+1 dimensional space-time, though not in 1+1dimension, l.u.b. may be shown to be non-associative contrary to the situation in Boolean algebra where union of sets *is* associative. (It is emphasised that the partial ordering and l.u.b. have been defined respectively for events and pairs of events and not for arbitrary subsets of all events).

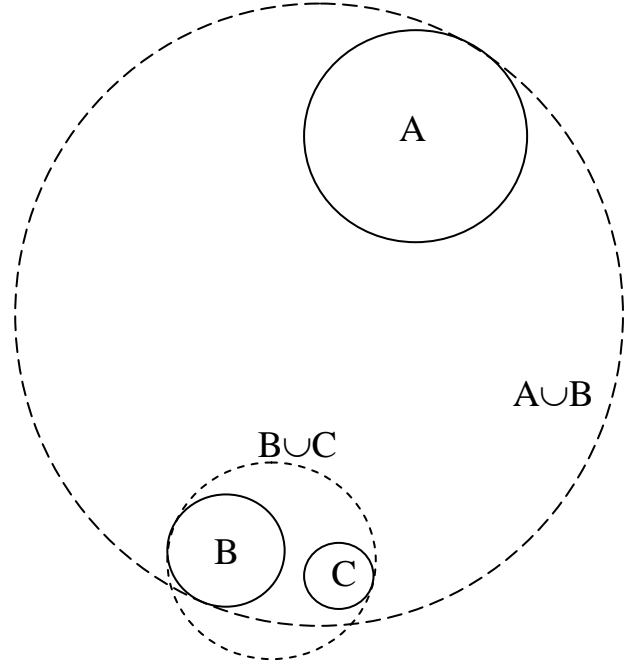

Figure 1 shows the intersection of the future light cones in 2+1 dimensional space-time of three events, A, B and C, with the plane $t = t_0$ for some observer. More recent events are represented by smaller light cone sections in the figure with remoteness in time proportional to the radius of the circle.

Proof that $\cup$ is non-associative:

In figure 1 $\quad A \cup B = (A \cup B) \cup C = A \cup (B \cup C) \geq B \cup C \quad$ if $\cup$ is associative

$\quad \therefore \quad \cup$ is non-associative by reduction ad absurdum

One could define $A \cup B \cup C$ to be the most recent event the future light cone of which contains A and B and C, but a new law of composition would then be needed for the l.u.b. of each subset of two or more events. It is a moot point as to whether the poset with l.u.b. defined for each pair of events just introduced may be regarded as a semi-lattice [14,15]. If so, it is a semilattice that is non-associative. Nevertheless, a meagre algebraic allowance of commutativity and idempotency of $\cup$ is sufficient for what is to follow.

It is now possible to show that a PT invariant association can be made between oriented spatial simplexes and oriented time displacements, though only in the absence of linear boosts as space-time symmetries. O is an event in Minkowski space-time and A, B and C are neighbouring mutually space-like separated events chosen to be simultaneous with O in some inertial frame. The spatial displacements from O to A, B and C are **a**, **b** and **c** and are entities with attitude, magnitude and orientation. OABC defines a simplex which has the same attributes. There is a family of such objects related by the translations and spatial rotations of the restricted form of the Poincaré group appropriate to the crystal class in question (the proper Poincaré group depleted of linear boosts and of all but a finite set of static rotations). A sign (+ or -) can be attached to each of the two orientations of the simplex OABC by forming the triple scalar product **a.b×c** provided only that an ordering of **a, b** and **c** to within cyclic permutation can be established. In the restricted version of the Poincaré group appropriate to crystal optics such an ordering is provided by the arrangement of **a**, **b** and **c** according to magnitude. Note again that by virtue of Lorentz contraction such an ordering is precluded when linear boosts are allowed space-time symmetries.

The sign of the orientation of the 3-space simplex is connected with ordering in time in the following way. The magnitudes of **a**, **b** and **c** are the spatial separations of pairs of simultaneous events that have l.u.b.s as already discussed. Furthermore **a**, **b** and **c** not being tied to any particular spatial location, their l.u.b.s can be brought by spatial translation to the same 3-space position and their future light cones can be nested somewhat in the manner of Russian dolls. Thus, the orientation of the spatial simplex is associated with an order relation in time. This association is invariant under space-time inversion PT. The fact that the association is not maintained under P and T separately means that these symmetries are less significant in crystal optics than elsewhere [16,17].

The incorporation of a vector product in the triple scalar product and its connection with time ordering means that the polar/axial and i/c characterization of tensors acquires a new level of subtlety. The governing PT symmetry is less restrictive of allowed phenomena than the separately applied P and T symmetries that work well in the analysis of static effects [5]. Specifically, fields and their response functions, and terms in equations in general, will be either PT-even or PT-odd, the former having axial-c as well as polar-i parts, the latter both polar-c and axial-i parts. This follows the general precept that all terms be allowed in dynamical equations apart from those specifically excluded on symmetry grounds.